# Absorption Spectra, Defect Site Distribution and Upconversion Excitation Spectra of CaF$_2$/SrF$_2$/BaF$_2$:Yb$^{3+}$:Er$^{3+}$ Nanoparticles

Sangeetha Balabhadra[1,2], Michael F. Reid[1,2], Vladimir Golovko[1,2], Jon-Paul R. Wells[1,2] *

[1]Dodd-Walls Centre for Photonic and Quantum Technologies, New Zealand

[2]School of Physical and Chemical Sciences, University of Canterbury, PB 4800, Christchurch 8140, New Zealand

* Corresponding author email: jon-paul.wells@canterbury.ac.nz

**HIGHLIGHTS**

- Size tuned cubic phase single ion doped (Yb$^{3+}$) and multi-ion doped (Yb$^{3+}$:Er$^{3+}$) MF$_2$ (M = Ba$^{2+}$, Sr$^{2+}$, and Ca$^{2+}$) nanoparticles were prepared by hydrothermal method.

- We report on the site distribution of Yb$^{3+}$ ions such as cubic (O$_h$) and cluster centres in Yb$^{3+}$:Er$^{3+}$ co-doped upconverting nanoparticles.

- The behaviour of Yb$^{3+}$ cluster centres was studied by tuning the metal cation in the alkaline earth fluoride (MF$_2$) host lattice from Ca$^{2+}$→Sr$^{2+}$→Ba$^{2+}$.

- Upconversion excitation spectra were obtained for the heterogeneous Yb-Er cluster centres, whilst scanning the excitation source through the wavelength dependent Yb$^{3+}$ absorption.

**ABSTRACT**

We report studies of the Yb$^{3+}$ site distribution in Er$^{3+}$ co-doped upconverting alkaline earth fluoride nanoparticles using high-resolution absorption measurements. It is found that Yb$^{3+}$ single ion cubic (O$_h$) sites and preferentially formed clusters are the dominant centres for moderate dopant concentrations. At higher concentrations, and for larger particle sizes, a minority C$_{4v}$(F$^-$) centre is also observed. Furthermore, the behaviour of Yb$^{3+}$ cluster centres, whose primary absorption band is centered at 10205 cm$^{-1}$ (in near resonance with 980 nm laser diodes) was studied by tuning the metal cation in the host lattice from Ca$^{2+}$→Sr$^{2+}$→Ba$^{2+}$. For co-doped materials, with heterogeneous cluster centres, $Er^{3+}$ upconversion fluorescence was observed upon selective excitation of $Yb^{3+}$, resulting in a strong visible emission from $^2H_{11/2}$, $^4S_{3/2}$ and $^4F_{9/2}$ multiplets. By monitoring the $Er^{3+}$ fluorescence, whilst scanning a laser through the wavelength dependent $Yb^{3+}$ absorption, we obtain an excitation spectrum for these heterogeneous cluster centres. Spectral features associated with excited state absorption as well as significant deviation from the linear absorption spectra are observed.

**GRAPHICAL ABSTRACT**

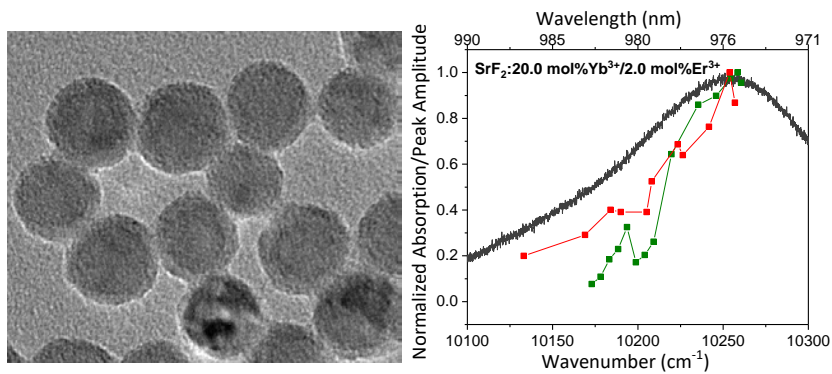

## 1. INTRODUCTION

Lanthanide ion ($Ln^{3+}$) doped upconverting nanoparticles are capable of converting low-energy near-infrared (NIR) photons into high-energy visible photons. This upconversion fluorescence has tremendous importance for a broad spectrum of applications in bio-imaging[1,2], drug delivery[3,4], theranostics[5], photovoltaics[6,7], security[8] and several applications in nanomedicine[9]. In general, upconverting nanomaterials consist of an inert, low-phonon energy, crystalline host matrix (*e.g.* fluorides) doped with a combination of optically active lanthanide ions ($Ln^{3+}$). One ion species acts as an energy absorbing sensitizer (typically $Yb^{3+}$ or $Nd^{3+}$) whilst another ion serves as the optically reporting species (*e.g.* $Er^{3+}$, $Tm^{3+}$, $Ho^{3+}$), which yields fluorescence at the desired wavelength[10]. Amongst the rare earth ions, $Yb^{3+}$ has a simple energy level structure with two multiplets ($^2F_{5/2}$, $^2F_{7/2}$) and large absorption cross-section near 980 nm, suitable for optical pumping at convenient semiconductor laser diode wavelengths. Despite promising applications in many technologically relevant areas, the comparatively low upconversion fluorescence yield typically obtained, limits practical application of such materials. Nanoparticles of varying size, dopant concentration, surface modifications, different host, and core/shell nanostructures have been explored to enhance the upconversion fluorescence quantum yield[11]. However, the authors are not aware of investigations which relate these variables to the absorption and excitation spectra.

It is crucial to understand the absorption and excitation dynamics of the of $Yb^{3+}$ ions in upconverting nanoparticles to determine the optimum pump wavelength and to gain insight into the energy transfer processes, which essentially govern the optical properties of lanthanide ion doped nanoparticles.

In the present work, $Ln^{3+}$ doped alkaline earth fluorides ($MF_2$, M=$Ca^{2+}$, $Sr^{2+}$, $Ba^{2+}$) were chosen as host materials owing to their well-known optical properties such as wide wavelength transparency, low phonon energy (495 cm$^{-1}$ for $CaF_2$, 366 cm$^{-1}$ for $SrF_2$, and 319 cm$^{-1}$ for $BaF_2$, respectively) and an extremely large band gap (12 eV)[12-14]. $MF_2$ crystals have a high-symmetry, cubic structure which is retained even after significant doping. In the cubic centre ($O_h$) the $F^-$ ions form a cage with an $F^-$ ion positioned non-locally at each corner in which the single $Yb^{3+}$ resides at the centre of each alternate cage. Incorporation of $Yb^{3+}$ occurs via the substitution of a cation ($M^{2+}$) requiring charge compensation via negative fluorine ions ($F^-$), positioned locally giving rise to trigonal ($C_{3v}(F^-)$) and tetragonal ($C_{4v}(F^-)$) $Yb^{3+}$ single ion centres [15-17]. However, at higher dopant concentrations preferential clustering of lanthanide ions occurs. It is suggested that these preferential cluster centres scavenge fluoride interstitials and as a result are negatively charged. Such cluster centres appear to have dimer, trimer and even hexameric configurations[18-20]. Numerous techniques such as NMR[21], EXAFS[22], EPR[13,23] and luminescence spectroscopy[12, 16, 24] have been employed for $Ln^{3+}$ site-distribution measurements (including the cluster sites), but predominantly for single crystals[25-27] and transparent ceramics[24, 28, 29]. We are not aware of similarly detailed studies of $Ln^{3+}$ doped nanoparticles.

We have synthesized water-dispersible, uniform $Ln^{3+}$ doped $MF_2$ nanoparticles by a simple hydrothermal method. The nanoparticle phase and structural identification has been performed using powder X-ray diffraction (PXRD) and transmission electron microscopy (TEM) techniques. The absorption spectra of as-synthesized nanoparticles was performed using a high resolution Fourier transform infra-red (FTIR) spectrometer. The nanoparticle site distribution was studied in the 10000-11000 cm$^{-1}$ spectral region and directly compared with bulk crystals of the same nominal concentration. The effect of $Yb^{3+}$ dopant concentration and nanoparticle size as well as the influence of variation in the ionic radius of the metal cations were systematically investigated. The $Er^{3+}$ upconversion fluorescence was recorded as a function of excitation wavelength around 980 nm to determine the sub-group of $Yb^{3+}$ ions, which most efficiently contribute to $Er^{3+}$ upconversion fluorescence via energy transfer.

## 2. EXPERIMENTAL

### 2.1. Materials

Anhydrous $CaCl_2$ (≥95.0%), $YbCl_3$ (K&K Laboratories Inc.), anhydrous $YbF_3$ (Alfa Aesar REacton, 99.9 %) and $CaF_2$ crystal offcuts (Crystran Ltd. UK) were used as starting materials. Ammonium fluoride ($NH_4F$, Aldrich 99.99 %) and sodium citrate (Fischer Scientific, 99.0 %) were used as the fluoride source and stabilizer. PbF$_2$ (Aldrich, 99 +%) was used as an oxygen scavenger. All the chemicals were used as received without further purification. MilliQ water, and acetone (Analytical grade, 99.98 %) were employed in all experiments.

## 2.2. Methods

### 2.2.1. Synthesis of upconverting nanoparticles

Citrate capped CaF$_2$ nanoparticles with different Yb$^{3+}$ molar concentrations (0.1, 1.0, 5.0 and 20.0 mol% in the synthesis medium) were prepared using a hydrothermal method[1]. In detail, stoichiometric quantities of anhydrous CaCl$_2$, and YbCl$_3$ to have nominal molar ratio of 0.99:0.01-Ca$^{2+}$:Yb$^{3+}$, with 3.5×10$^{−3}$ mol of the total metal amount, were dissolved in 5 mL of deionized water in a Teflon vessel. A sodium citrate solution (20 mL of a 1.0 M) was added under vigorous stirring for 5 minutes. Then an ammonium fluoride aqueous solution (2.5 mL of a 3.5 M ) was added to the reaction mixture, in order to have a slight excess of fluoride ions, (M+Ln$^{3+}$):F=1:2.5 molar ratios. The clear solution obtained was heat treated at 190°C for 6 hours under autogenous pressure in an autoclave. After that, the autoclave was quickly cooled down to room temperature in order to quench the reaction and to limit further growth (in size) of the nanoparticles. The resulting nanoparticles were obtained by centrifugation. Later washed with milliQ water and acetone, and collected by centrifugation (8000 rpm, 10 minutes). Nanoparticles are easily dispersible in water to form transparent colloids. To obtain nanoparticles in powder form, the gel is dried at room temperature for 24 hours in air. The samples were denoted according to their molar concentrations of dopants used in the synthesis. Larger nanoparticles were prepared in the same procedure by increasing the reaction time to 48 hours and 96 hours, respectively. Similarly, the Yb$^{3+}$/Er$^{3+}$ co-doped upconverting MF$_2$ (M= Ca$^{2+}$, Sr$^{2+}$, Ba$^{2+}$) nanoparticles were prepared by using optimal molar concentrations of Yb$^{3+}$/Er$^{3+}$:20/2, in order to avoid concentration quenching and the corresponding non-radiative relaxation processes.

### 2.2.2. Synthesis of CaF$_2$:Yb$^{3+}$ bulk crystals

In order to have a direct comparison, bulk CaF$_2$ crystals were prepared with same Yb$^{3+}$ nominal concentrations (0.1, 1.0, 5.0 and 20.0 mol%) as nanoparticles. The crystals were grown in graphite crucibles under high vacuum (10$^{−5}$ Torr), using the vertical Bridgman technique in a 38 kW radio frequency furnace, at temperatures up to 1500˚C[2]. The starting material was prepared by crushing up the offcuts of CaF$_2$ and mixing the appropriate amounts of YbF$_3$. A small amount of PbF$_2$ was added to the charge in order to act as an oxygen scavenger, also resulting in a crystal with reduced concentration of the divalent species. Annealing is achieved via an automatic cooling cycle which leads to improved quality crystals. The as-grown crystals were then cut using a Struers Minitom diamond saw and polished using a Struers LaboPol-2 to give an excellent surface finish for absorption measurements.

## 2.3. Experimental details
### 2.3.1. Powder X-ray diffraction

Phase identification of the as synthesized powder samples was inferred from their X-ray diffraction patterns. Corresponding diffractograms were collected on a Rigaku SmartLab X-ray diffractometer with CuKα1 radiation, 1.5406 Å, operating at 40 kV and 30 mA, in the $2\theta$ range 20° to 90° with a 0.01° step size in the reflection scanning mode. The reference data were taken from the International Centre for Diffraction Data (ICDD) database.

### 2.3.2. Transmission Electron Microscopy

The morphology of the as synthesized nanoparticle suspensions was analysed on a Philips-CM200 transmission electron microscope (TEM) connected to Gatan Orus CCD camera, operated at 200 kV.

### 2.3.3. Dynamic Light Scattering

Dynamic light scattering (DLS) measurements were carried out using a Malvern Zetasizer Nano ZS instrument operating at 532 nm with a 50 mW laser. The data were acquired for sodium citrate capped $CaF_2:Yb^{3+}$ nanoparticles dispersed to have 0.25 wt% aqueous suspension. The reported values are the average of three measurements.

### 2.3.4. Fourier-Transform Infrared Spectroscopy (FTIR)

Temperature dependent infrared absorption measurements were performed using Bruker Vertex 80 FTIR having 0.075 cm$^{-1}$ resolution with an optical path purged by $N_2$ gas.

For the absorption measurements, the air-dried powders of nanoparticles were pressed using a pellet maker in order to form a thin pellet (~1 mm) which was then placed into a small copper sample holder. In the case of crystals, the boule was prepared by first cutting either end with a diamond saw and the ends were polished by different grades of sandpaper. As polished boule was inserted into the copper holder for samples with low $Yb^{3+}$ concentrations and for the high concentration $Yb^{3+}$ samples, thin slices (~1 mm) were cut off of the end of the boule and polished in the same manner as the boule. The samples were cooled in closed-cycle helium cryostat with temperature variability provided by a resistive heating element.

### 2.3.5. Diode laser emission spectroscopy

Temperature dependent upconversion emission spectra were measured using an Intense 3120 model laser diode controlled with NewPort 560B laser diode driver and NewPort 350B temperature controller. The laser wavelength is tuned from 975 to 982 nm by changing the laser diode temperature between 283 and 293 K. The free space propagating laser beam was focused on the sample using a LA1608 plano-convex ($f$ = 75.0 mm) aspheric lens (Thorlabs). The laser output power is measured with FieldmaxII laser power meter coupled with the Coherent Powermax (Model PM10) sensor. A range of 0.1−4.0 ND filters (Thorlabs) were used to maintain the laser power at a constant value throughout an experiment. The emission spectra in the visible region was recorded on a modular double grating iHR550 spectrophotometer (Horiba) coupled to a H9170

Hamamatsu photomultiplier. The samples were cooled in Janis closed-cycle helium cryostat with temperature variability provided by a resistive heater controlled by a LakeShore 325 temperature controller. An external UV-VIS mini-spectrometer (Ocean Optics P1000 UV-VIS) was used to monitor the laser diode wavelength.

## 3. RESULTS AND DISCUSSION

Fig. 1a shows the powder X-ray diffraction patterns of $CaF_2:Yb^{3+}$ nanoparticles for varying reaction times from 6 to 96 hours ((i)-(iii) at a constant $Yb^{3+}$ molar concentration of 1.0 mol%) and for different concentrations of $Yb^{3+}$ in the synthesis media between 0.1 and 20.0 mol% ((iv)-(vi) using a constant reaction time of 6 hours). The samples exhibit the cubic phase of $CaF_2$ (space group $Fm\overline{3}m$), in agreement with ICDD card (04−002−4443) and as reported by Pedroni et al.[30]. No new diffraction peaks are observed when the amount of $Yb^{3+}$ increases from 0.1 to 20.0 mol% nor as the reaction time increases, indicating that these ions have been effectively introduced into the $CaF_2$ lattice. The diffraction peaks of the nanoparticles in Fig. 1a ((iii) and (vi)) show a slight shift towards lower angles, in comparison to the other nanoparticles which can be due to the sample displacement.

Fig. 1b displays the powder X-ray diffraction patterns of $MF_2$:20.0 mol% $Yb^{3+}$:2.0 mol% $Er^{3+}$ nanoparticles where M=$Ba^{2+}$, $Sr^{2+}$, and $Ca^{2+}$ ((i)-(iii)), respectively. The samples are of pure cubic crystalline phase (space group $Fm\overline{3}m$), in agreement with ICDD cards 00-004-0452 for $BaF_2$, 04-016-6744 for $SrF_2$, and 04-002-4443 for $CaF_2$, respectively. From the diffractograms the average size of the crystallite domains in nanoparticles could be estimated using Scherrer's equation (Table S1 in supporting information) using the full-width half-maximum value of the $2\theta$ diffraction peak at 46.9° (i.e. the (220) plane). In addition, the hydrodynamic size distribution of the nanoparticles was also measured using dynamic light scattering (DLS, see Fig. S1 and Table S1 in the supporting information). The hydrodynamic sizes are in good agreement with the average crystallite sizes calculated from the PXRD results. Representative transmission electron microscopy (TEM) images of the $CaF_2$:5.0 mol% $Yb^{3+}$ nanoparticles are shown in Fig. 1c which illustrates that the nanoparticles are spherical and homogeneously distributed. As shown in the size distribution histogram in Fig. 1d, the diameter of the nanoparticles ranges from 8 to 22 nm, with average values of 15.8±2.1 nm. The TEM nanoparticle size distribution is in good agreement with the hydrodynamic sizes from DLS and the average crystallite sizes calculated from the PXRD results.

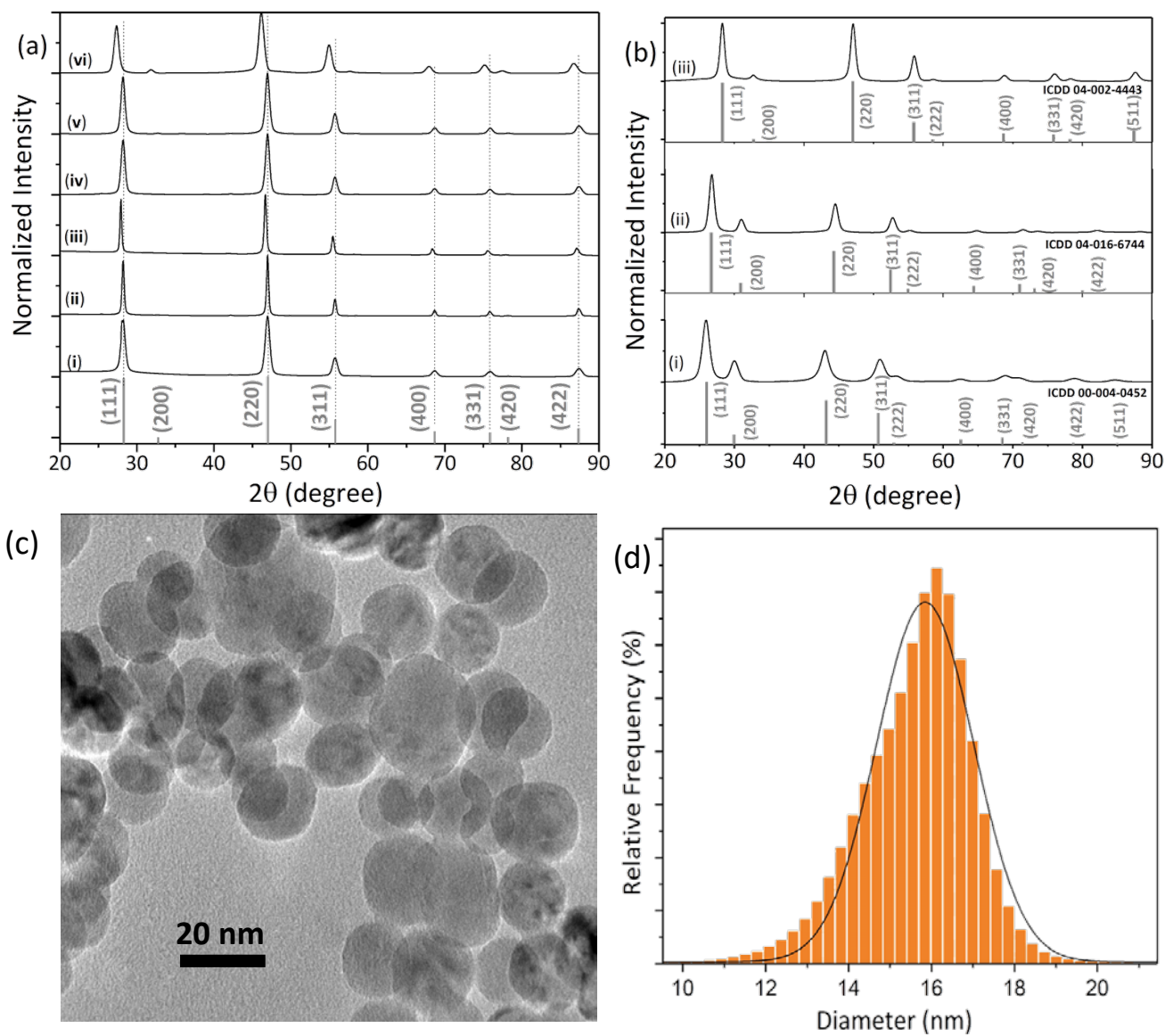

Fig. 1. Powder X-ray diffraction patterns of (a) CaF$_2$ nanoparticles prepared at reaction times (i) 6 hours, (ii) 48 hours and (iii) 96 hours for 1 mol% of Yb$^{3+}$ molar concentrations; doped with (iv) 0.1 mol%, (v) 5.0 mol%, and (vi) 20.0 mol% of Yb$^{3+}$ concentrations at reaction time of 6 hours. (b) MF$_2$:20.0 mol% Yb$^{3+}$:2.0 mol% Er$^{3+}$ nanoparticles where (i) M=Ba$^{2+}$, (ii) M=Sr$^{2+}$, and (iii) M=Ca$^{2+}$. The diffraction patterns of cubic BaF$_2$ (ICDD Card No 00-004-0452), SrF$_2$ (ICDD Card No 04-016-6744), and CaF$_2$ (ICDD Card No 04-002-4443), are also depicted. (c) Transmission electron microscope image and (d) corresponding size distribution histogram of CaF$_2$:5.0 mol% Yb$^{3+}$ nanoparticles. A total of 100 nanoparticles were measured. The solid line is the best fit of the experimental data to a Gaussian distribution ($r^2 > 0.989$).

Fig. 2a and b show the FTIR absorption spectra measured for nanoparticles and bulk crystals as a function of Yb$^{3+}$ ion molar concentration, recorded for samples cooled to 10 K (at 25 K for the CaF$_2$:0.1 mol% Yb$^{3+}$ crystal). All samples display characteristic absorption features associated with the $^2F_{7/2} \rightarrow {}^2F_{5/2}$ Yb$^{3+}$ inter-multiplet transition in the 10100−11000 cm$^{-1}$ spectral region. These absorption spectra have been extensively studied for bulk crystals, and the literature data[12, 16, 17, 31] is in excellent agreement with our observations for high quality, oxygen free samples. The CaF$_2$:0.1 mol% Yb$^{3+}$ crystal (Fig. 2b (i)) has very sharp lines, corresponding to the single ion centres: cubic (O$_h$:10381 and 10840 cm$^{-1}$), tetragonal (C$_{4v}$(F$^-$):10321, 10401 and 11000 cm$^{-1}$), trigonal (C$_{3v}$(F$^-$):10257, 10375, and 10712 cm$^{-1}$) as well as lines assigned to Yb$^{3+}$ clusters (10199, and 10205 cm$^{-1}$). In crystals with high Yb$^{3+}$ concentrations (Fig. 2b ((ii)-(iv))) the absorption spectra are completely dominated by cubic and clusters centres. It can be speculated that as the population of negatively charged cluster centres increases, there is an associated rise in the population of non-locally charge compensated cubic centres.

For nanoparticles doped with 0.1 mol% Yb$^{3+}$ the absorption spectra consist of isolated cubic centres (corresponding to the feature at 10381 cm$^{-1}$). The feature denoted by an * appears to be associated with perturbed cubic centres, possibly residing near the nanoparticle surface. We note that the bulk crystal are synthesised by the Bridgman-Stockbarger technique (1500˚C) and the nanoparticles by a hydrothermal method (190°C) (refer supporting information for details of the corresponding synthesis methods)[32-34]. High-temperature crystal growth gives scope for the ions to freely move and to arrange themselves in a minimum energy configuration[35]. This favours the formation of lower symmetry sites such as the C$_{3v}$(F$^-$) and C$_{4v}$(F$^-$) centres. At higher Yb$^{3+}$ concentrations, the spectra indicate that the nanoparticle site distribution resembles more closely that observed for the bulk crystals - being dominated by cluster and cubic sites. However, there are clearly differences in detail. Both in bulk crystals and nanoparticles, the feature centred around 10850 cm$^{-1}$ contains contributions from both the cubic and cluster centres. With an increase in Yb$^{3+}$ concentration there is a noticeable increase of this feature along with the cluster centre absorption feature at 10205 cm$^{-1}$. The absorption spectra recorded at room temperature for both bulk crystals and nanoparticles (Fig. S2 in supporting information) clearly shows both the cubic and cluster centres, although the transitions are broadened in comparison with 10 K.

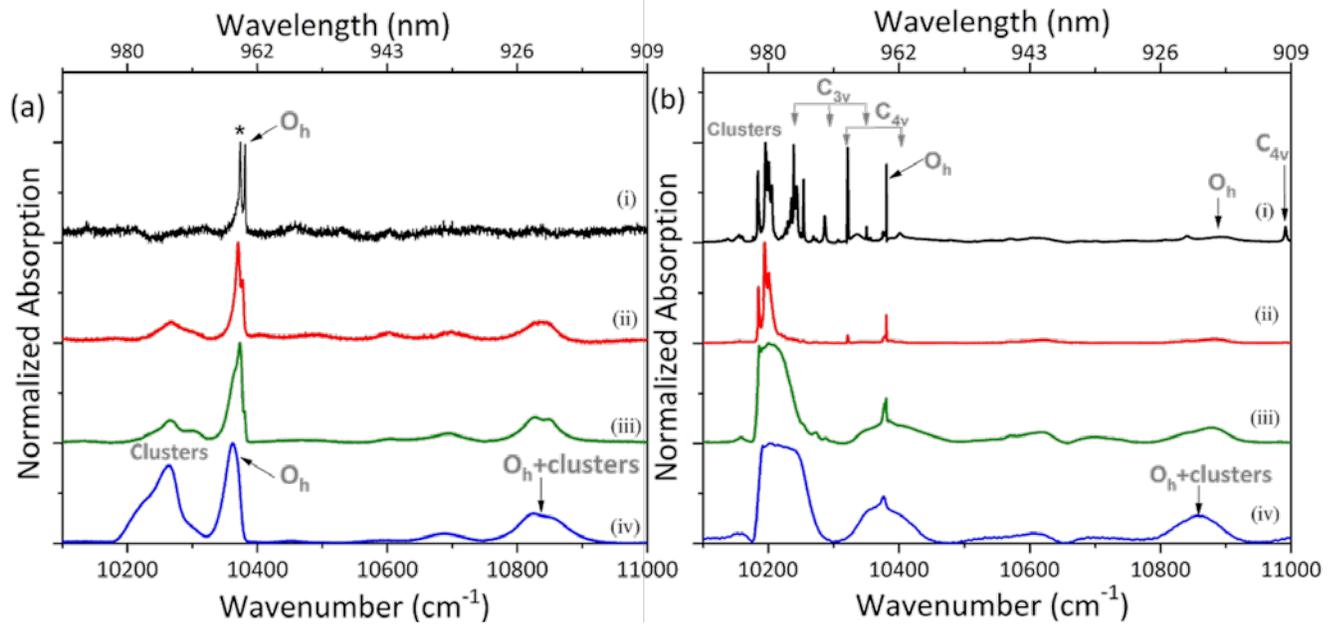

Fig. 2. FTIR absorption spectra measured at 10 K for CaF$_2$ doped with (i) 0.1, (ii) 1.0, (iii) 5.0 and (iv) 20.0 mol% Yb$^{3+}$ concentrations for (a) nanoparticles and (b) crystals. The spectrum for the CaF$_2$:0.1 mol% Yb$^{3+}$ crystal was recorded at 25 K. The feature marked * appears to be perturbed cubic centres.

To understand the influence of nanoparticle size on the observed site distribution, different reaction times between 6, 48 and 96 hours were used, resulting in particle diameters increasing from 12.1 nm, to 35.4 nm and to 52.6 nm, increasing the reaction time. The average diameters of the nanoparticles measured using TEM correlate nicely with crystallite sizes based on PXRD and hydrodynamic sizes based on DLS measurements (Fig. 1, Fig. S1 and Table S1 in supporting information). In the larger nanoparticles, the absorption spectra also have weak absorption features from tetragonal C$_{4v}$(F$^-$) sites (see Fig. 3).

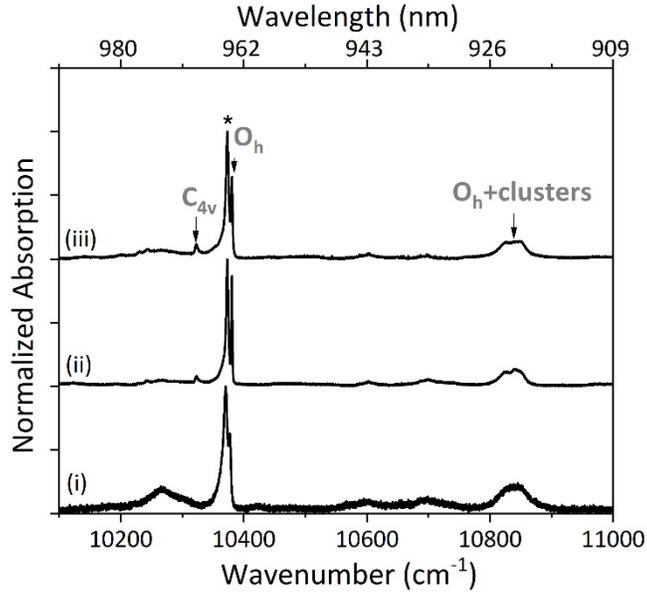

Fig. 3. FTIR absorption spectra measured at 10 K for CaF$_2$:1.0 mol% Yb$^{3+}$ nanoparticles at reaction times of (i) 6 hours, (ii) 48 hours and (iii) 96 hours, respectively. The feature marked * appears to be perturbed cubic centres.

Upconverting nanoparticles containing Yb$^{3+}$ as the energy absorbing ion, are commonly pumped by strained layer, multi-quantum well InGaAs laser diodes having an output wavelength near 980 nm (10205 cm$^{-1}$). The typical dopant concentrations used lie in the 16.0-20.0 mol% range due to the requirement for high absorption depth (see Fig. 2a (iv))[36]. It would be standard for the reporting ion (*e.g.* Er$^{3+}$) to be doped in the 1.0-2.0 mol% range.

Fig. 4((i)-(iii)) shows the 10 K FTIR absorption spectra for 20.0 mol% Yb$^{3+}$: 20.0 mol% Er$^{3+}$ and 20.0 mol% Yb$^{3+}$:2.0 mol% Er$^{3+}$ co-doped upconverting nanoparticles. The absorption spectrum of 20.0 mol% Yb$^{3+}$ doped CaF$_2$ nanoparticles shown in Fig. 4(i) has features attributable to cubic as well as cluster centres. When doped with Er$^{3+}$, the prominent characteristics of the Yb$^{3+}$ single ion sites remain essentially intact. Trivalent erbium has a broad absorption band which resides near 10340 cm$^{-1}$ corresponding to the $^4I_{15/2} \rightarrow {}^4I_{11/2}$ transition[37], whose absorption tail closely overlaps with the Yb$^{3+}$ cluster centre absorption. The absorption cross-section of Yb$^{3+}$ ions in the 10100−11000 cm$^{-1}$ region is much larger than that for the Er$^{3+}$ ions (Fig. 4) thus the energy transfer from Yb$^{3+}$ to Er$^{3+}$ via the channel $^2F_{5/2}(Yb^{3+}) + {}^4I_{15/2}(Er^{3+}) \rightarrow {}^2F_{7/2}(Yb^{3+}) + {}^4I_{11/2}(Er^{3+})$ efficiently sensitizes the luminescence of Er$^{3+}$[38].

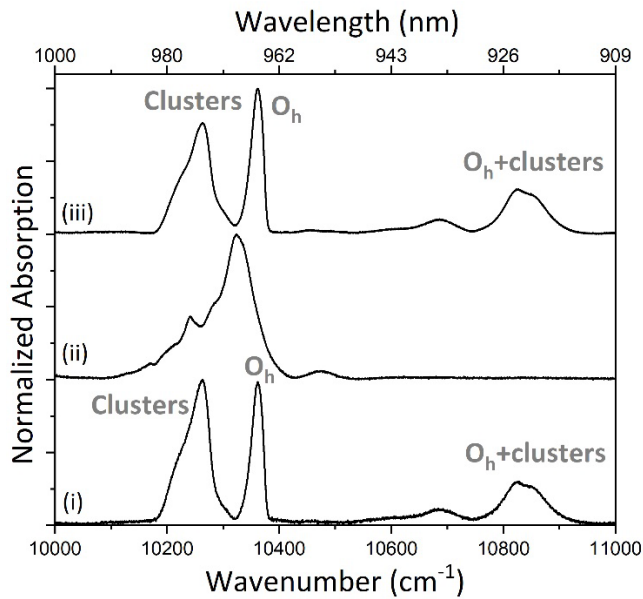

Fig. 4. FTIR absorption spectra measured at 10 K for $CaF_2$ nanoparticles doped with (i) 20.0 mol% $Yb^{3+}$, (ii) 20.0 mol% $Er^{3+}$ and (iii) 20.0 mol% $Yb^{3+}$/2.0 mol% $Er^{3+}$.

It becomes immediately clear that an excitation source fixed at 980 nm (having a typical linewidth of approximately 3 nm, *e.g.* an Intense model 3120 laser diode) is not optimally resonant with the absorption features of the $Yb^{3+}$ clusters in $CaF_2$ based nanoparticles. Varying the cation may yield a shift in the resonance peak thus improving the coupling efficiency. We have measured the absorption spectra of $Yb^{3+}$:$Er^{3+}$ co-doped $MF_2$ (M= $Ba^{2+}$, $Sr^{2+}$, and $Ca^{2+}$) nanoparticles which are shown in Fig. 5a. Changing the cation in the sequence from $Ca^{2+}\rightarrow Sr^{2+}\rightarrow Ba^{2+}$ results in a very significant broadening of the absorption transitions which is highly favorable as an energy conduit which is stable against minor wavelength fluctuations in the pump laser system. It also suggests broadband excitation via LEDs might be a viable option.

Fig. 5b shows the upconversion fluorescence spectra, excited using a 980 nm laser diode employing a power density of 31 W·cm$^{-2}$ at the sample. Application relevant room temperature fluorescence spectra are presented here (the 10 K data (Fig. S5) can be found in the supporting information). All three upconverting nanoparticle samples exhibit upconversion fluorescence which can be assigned to the $^2H_{11/2}\rightarrow{}^4I_{15/2}$ (520 nm), $^4S_{3/2}\rightarrow{}^4I_{15/2}$ (540 nm) and $^4F_{9/2}\rightarrow{}^4I_{15/2}$ (650 nm) transitions of $Er^{3+}$. The upconverting nanoparticle fluorescence follows same trend at cryogenic temperatures (Fig. S5). However by tuning the laser to 975 nm, the absorption is significantly higher for all three samples (when compared for nanoparticles with similar dopant concentration) and we observe an increase in the upconversion fluorescence intensities as shown in Fig. 5c. Overall, it is evident that $SrF_2$ upconverting nanoparticles exhibit the most intense upconversion fluorescence, which we speculate is due to the low phonon-energy (366 cm$^{-1}$) and a strong absorption band in the 980-970 nm region. We note that although $BaF_2$ has a lower phonon energy (319 cm$^{-1}$) it has very weak emission (Fig. 5b inset) due to its spectrally broad and low peak amplitude absorption spectra – these may be more suitably pumped with a broadband light source. Moreover, $BaF_2$ nanocrystals tend to have a distorted crystal structure when high concentrations of low ionic radius $Yb^{3+}$ (0.086 nm) and $Er^{3+}$ (0.089 nm) replace the high ionic radius $Ba^{2+}$ (0.135 nm). In $CaF_2$ nanoparticles, the upconversion fluorescence yields predominantly red emission ($^4F_{9/2} \rightarrow ^4I_{15/2}$) when compared to the observed spectra for the $SrF_2$ and $BaF_2$ nanoparticles. This is presumably a result of the high-phonon energy (495 cm$^{-1}$) giving a higher probability of non-radiative relaxation to the lower energy emitting state in the $CaF_2$ host lattice[38].

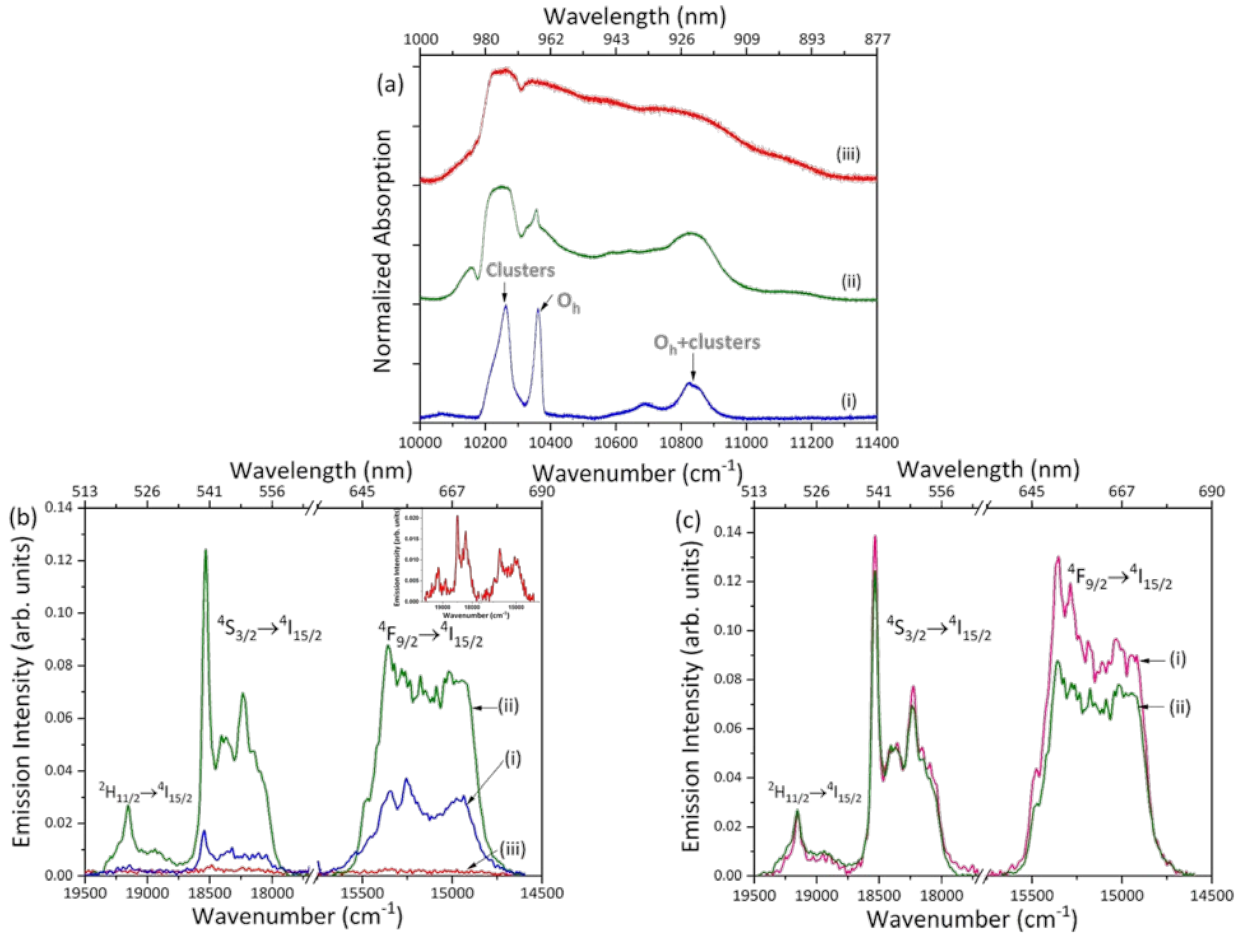

Fig. 5. (a) FTIR absorption spectra measured at 10 K and (b) room temperature upconversion fluorescence spectra measured for $MF_2$:20.0 mol% $Yb^{3+}$/2.0 mol% $Er^{3+}$ nanoparticles where M = (i) $Ca^{2+}$ (ii) $Sr^{2+}$ and (iii) $Ba^{2+}$ nanoparticles at 980 nm laser excitation and (c) $SrF_2$:20.0 mol% $Yb^{3+}$/2.0 mol% $Er^{3+}$ nanoparticles exciting at laser wavelength of (i) 975 nm and (ii) 980 nm. Inset in (b) is the zoom in of upconversion fluorescence spectrum of $BaF_2$:20.0 mol% $Yb^{3+}$/2.0 mol% $Er^{3+}$ nanoparticles on an expanded scale. A laser power density of 31 W·cm$^{-2}$ was used to record the upconversion fluorescence spectra.

Fig. 6 shows the excitation spectra obtained monitoring the $Er^{3+}$ fluorescence $^4S_{3/2} \rightarrow ^4I_{15/2}$ (540 nm, green dotted lines) and $^4F_{9/2} \rightarrow ^4I_{15/2}$ (650 nm, red dotted lines) transitions, while scanning the laser sequentially across the $Yb^{3+}$ absorption profile, measured for all three bare nanoparticle materials and compared against their corresponding absorption spectra (solid lines in Fig. 6). These excitation and absorption spectra were measured at room temperature. The excitation profile was recorded at low laser power densities 4 W·cm$^{-2}$ (SrF$_2$ and CaF$_2$) and 14 W·cm$^{-2}$ (BaF$_2$) where UC fluorescence has a linear dependence with a slope of approximately two when plotted on a log-log plot. Cryogenically cooling the samples yields no appreciable difference. Within the wavelength range of the diode used, all three materials exhibit the highest emission yield near 10250 cm$^{-1}$ (975 nm) which is also where the nanoparticles have the maximum absorption. However, the spectral dependence of the upconversion fluorescence *does not*, in all cases, *reflect the absorption spectra*. The observed deviation becomes successively more pronounced as the ionic radius of the metal cation increases. For BaF$_2$ nanoparticles the upconversion fluorescence amplitude decreases to 30% of its peak value at 10150 cm$^{-1}$, however the absorption depth at this frequency is unchanged. It is also observable that a comparatively sharp peak is present at 10194 and 10215 cm$^{-1}$ for SrF$_2$ and BaF$_2$ respectively. Again, this is not reflected in the absorption spectra and appears to be an excited state absorption feature. Of course, upconversion is a complex, nonlinear, process, but we may tentatively interpret the lack of simple proportionality as indicating that only a subset of the absorbing Yb$^{3+}$ ions are located in heterogeneous clusters contained the emitting ion, Er$^{3+}$, with the remaining ions residing in clusters containing only Yb$^{3+}$ ions (Fig. 6). Thus the Yb absorption profile cannot be used as an indicator of the optimum excitation wavelength and measurement of both the absorption and upconversion excitation spectra of upconverting nanomaterials (especially the most commonly used nanoparticles *e.g.* NaYF$_4$) is required, in order to increase the upconversion fluorescence yield, which is the bottleneck holding back practical implementation of the existing nanoparticle materials.

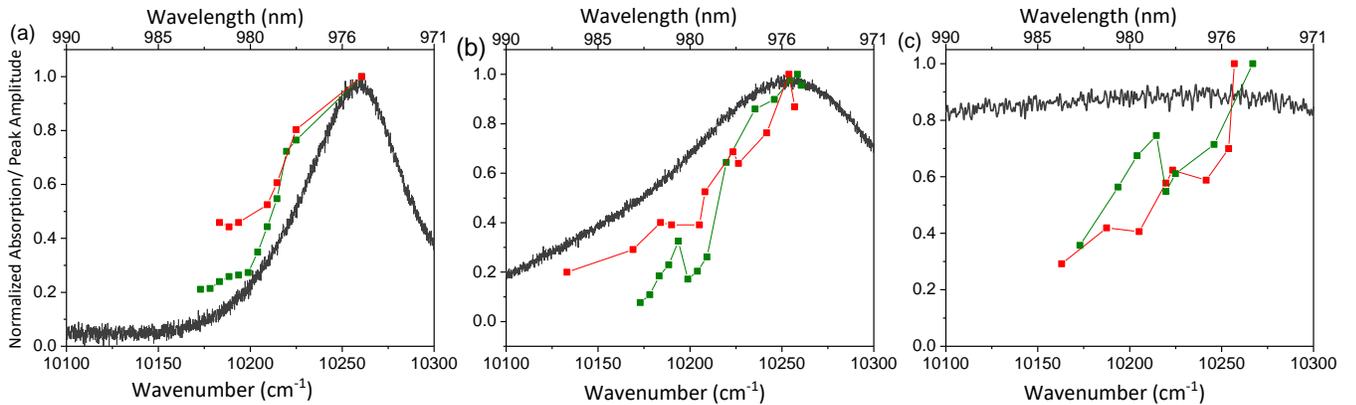

Fig. 6. Yb$^{3+}$ excitation profile measured monitoring Er$^{3+}$ ion $^4S_{3/2}\rightarrow{}^4I_{15/2}$ (540 nm) peak amplitude for MF$_2$:20.0 mol%Yb$^{3+}$/2.0 mol% Er$^{3+}$ nanoparticles where M = (a) Ca$^{2+}$ (b) Sr$^{2+}$ and (c) Ba$^{2+}$ nanoparticles at room temperature, respectively. The excitation profiles were measured for the Er$^{3+}$ $^4S_{3/2}\rightarrow{}^4I_{15/2}$ (540 nm, green dotted lines) and $^4F_{9/2}\rightarrow{}^4I_{15/2}$ (650 nm, red dotted lines) transitions. All data have been normalized to the maximum data point.

## 4. CONCLUSIONS

In summary, cubic phase single ion doped (Yb$^{3+}$) and multi-ion doped (Yb$^{3+}$/Er$^{3+}$) MF$_2$ (M = Ca$^{2+}$, Sr$^{2+}$, and Ba$^{2+}$) nanoparticles have been successfully synthesized by a simple hydrothermal route at moderate temperatures and ambient pressure. Their structure and morphology were characterized by the PXRD, TEM, and DLS techniques. The absorption spectra of the as synthesized nanoparticles were investigated using FTIR spectroscopy and directly compared with data for the bulk single crystals of the same dopant concentrations. The $Yb^{3+}$ ion site distribution is dominated by both non-locally charge compensated cubic centres and preferentially formed multi-ion cluster centres. Measurements of the spectral dependence of the $Er^{3+}$ upconversion fluorescence obtained whilst pumping $Yb^{3+}$ exhibit increasingly large deviations from the linear absorption spectrum as the size of the metal cation increases. In addition, excited state absorption features can be observed. The former appears to indicate an increase in homogeneous Yb-Yb clustering however the absorption maximum is still observed to coincide with the maximum in the $Er^{3+}$ upconversion fluorescence.

**CRediT AUTHORSHIP CONTRIBUTION STATEMENT**

**Sangeetha Balabhadra:** Synthesis and experimental work, data analysis, writing, editing

**Michael Reid:** Data analysis, and reviewing

**Vladimir Golovko:** Manuscript revision

**Jon-Paul Wells:** Assisting in spectroscopic measurements, data analysis, writing and editing

**DECLARATION OF COMPETING INTERESTS**

The authors declare that they have no known competing financial interests or personal relationships that could have appeared to influence the work reported in this paper.

**APPENDIX A. SUPPLEMENTARY DATA**

The supplementary data contains results of DLS measurements, absorption and fluorescence spectra.

**ACKNOWLEDGEMENTS**

The authors acknowledge the expert technical assistance of Dr Mathew Polson, Mr Shaun Mucalo, Mr Stephen Hemmingsen, Mr Robert Thirkettle and Mr Nick Olivier.

# Supporting Information

# Absorption Spectra, Defect Site Distribution and Upconversion Excitation Spectra of CaF$_2$/SrF$_2$/BaF$_2$:Yb$^{3+}$:Er$^{3+}$ Nanoparticles

Sangeetha Balabhadra[1,2], Michael F. Reid[1,2], Vladimir Golovko[1,2], Jon-Paul R. Wells[1,2] *

[1]Dodd-Walls Centre for Photonic and Quantum Technologies, New Zealand

[2]School of Physical and Chemical Sciences, University of Canterbury, PB 4800, Christchurch 8140, New Zealand

* Corresponding author email: jon-paul.wells@canterbury.ac.nz

Table S1. Summary of synthesized samples, reaction time (in hours) of the synthesis, average crystallite sizes calculated from PXRD diffractograms and average hydrodynamic size of the nanoparticles measured in DLS.

| S.No | Sample | Reaction time (in hours) | Average crystallite sizes (nm from PXRD) | Average NPs sizes (nm from DLS) |
|---|---|---|---|---|
| 1 | $CaF_2$:0.1 mol% $Yb^{3+}$ | 6 | 13.2 | 14.3 |
| 2 | $CaF_2$:1 mol% $Yb^{3+}$ | 6 | 12.1 | 16.5 |
| 3 | $CaF_2$:5 mol% $Yb^{3+}$ | 6 | 12.6 | 20.2 |
| 4 | $CaF_2$:20 mol% $Yb^{3+}$ | 6 | 11.2 | 22.3 |
| 5 | $CaF_2$:1 mol% $Yb^{3+}$ | 24 | 26.4 | 35.4 |
| 6 | $CaF_2$:1 mol% $Yb^{3+}$ | 48 | 52.6 | 63.2 |
| 7 | $CaF_2$:20 mol% $Yb^{3+}$ | 6 | 14.1 | 18.6 |
| 8 | $CaF_2$:20 mol% $Yb^{3+}$/2 mol% $Er^{3+}$ | 6 | 10.9 | 13.2 |
| 9 | $SrF_2$:20 mol% $Yb^{3+}$/2 mol% $Er^{3+}$ | 6 | 10.6 | 15.8 |
| 10 | $BaF_2$:20 mol% $Yb^{3+}$/2 mol% $Er^{3+}$ | 6 | 6.4 | 12.4 |

## 1. RESULTS

### 1.1. Dynamic Light Scattering

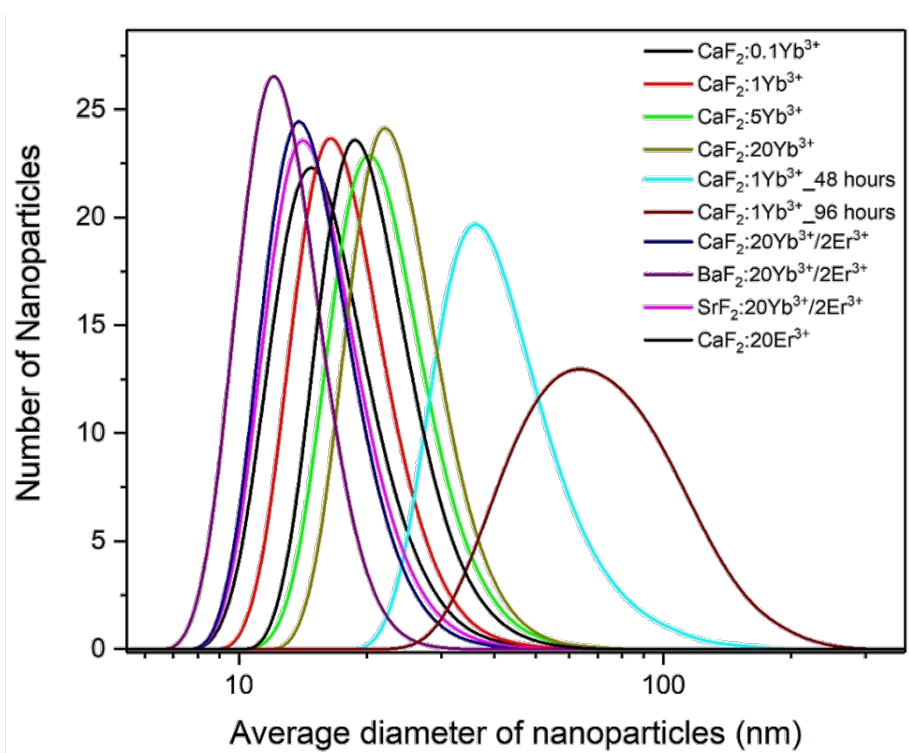

Fig. S1. Dynamic light scattering measurements of all the studied nanoparticles. \

## 1.2. Room temperature absorption spectra

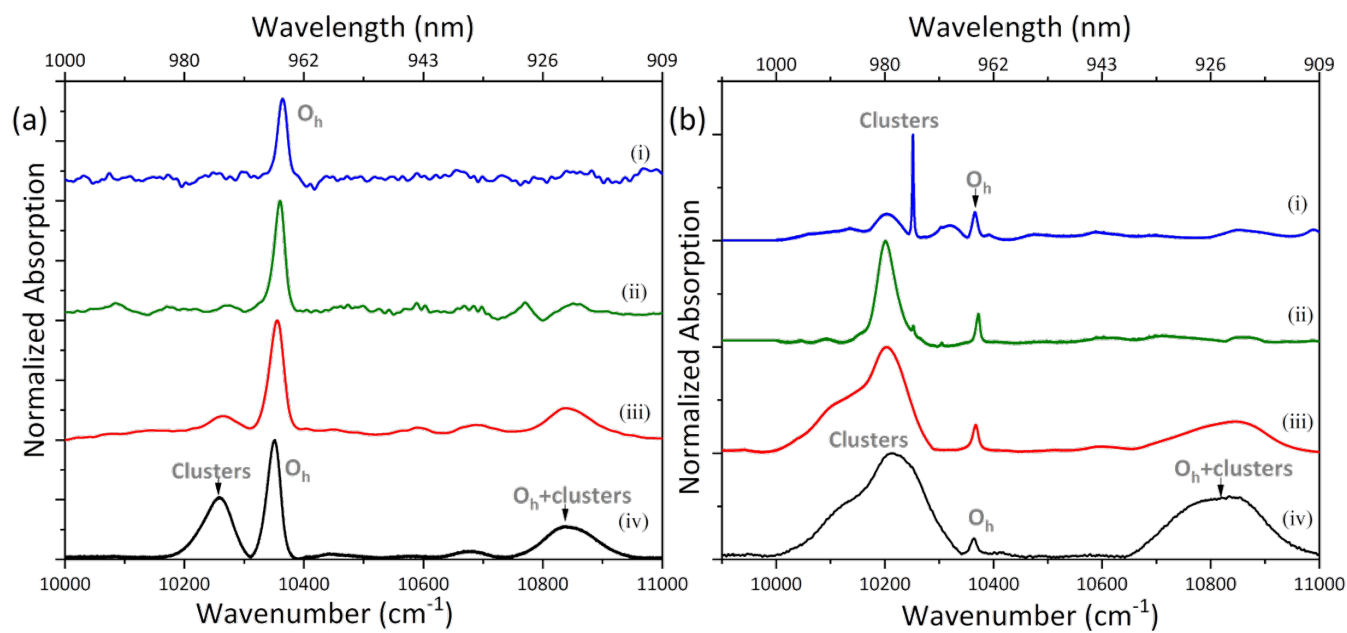

Fig. S2. Room temperature FTIR absorption spectra of CaF$_2$ for (i) 0.1, (ii) 1.0, (iii) 5.0 and (iv) 20.0 mol% Yb$^{3+}$ concentrations for (a) nanoparticles and (b) crystals.

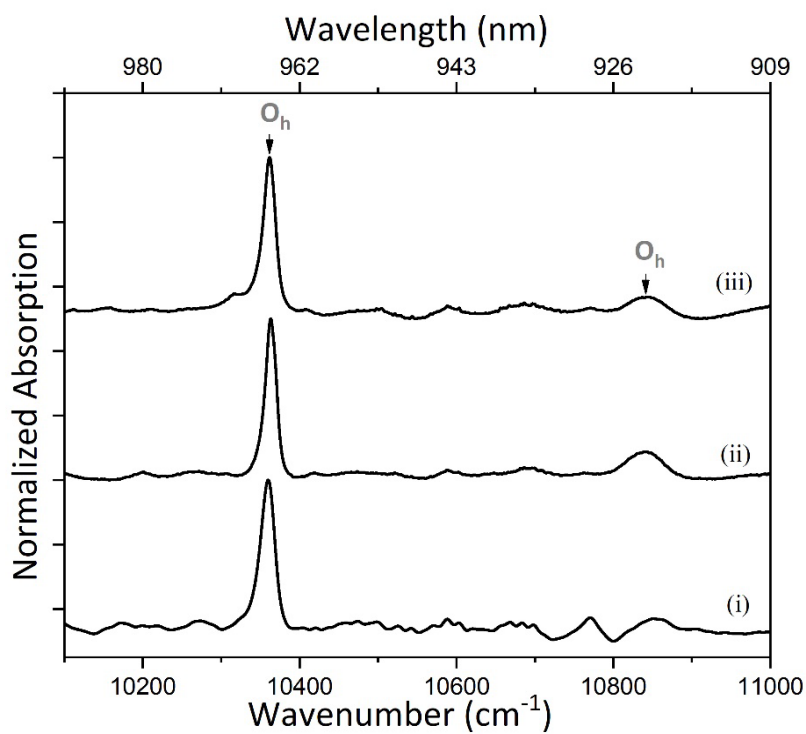

Fig. S3. Room temperature FTIR absorption spectra of CaF$_2$:1 mol% Yb$^{3+}$ nanoparticles synthesized using reaction times of (i) 6 hours, (ii) 48 hours and (iii) 96 hours, respectively.

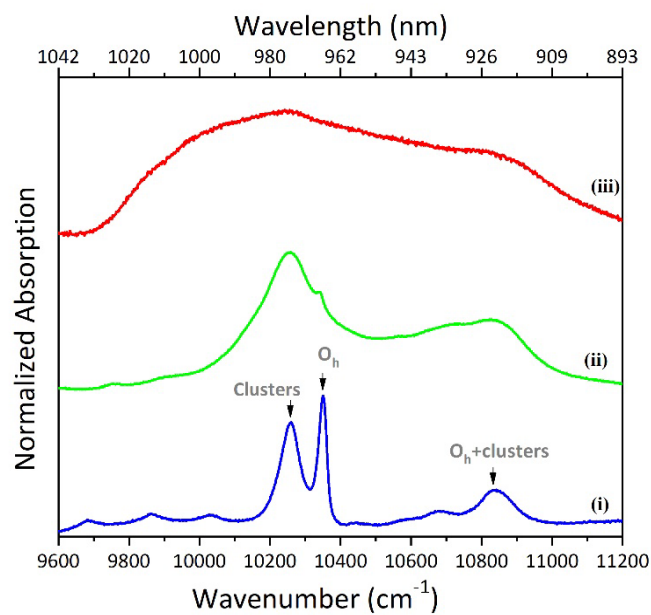

Fig. S4. Room temperature FTIR absorption spectra of $MF_2$:20 mol% $Yb^{3+}$/2 mol% $Er^{3+}$ nanoparticles where M=( i) $Ca^{2+}$, (ii) $Sr^{2+}$ and (iii) $Ba^{2+}$, respectively.

### 1.3. Upconversion fluorescence spectra

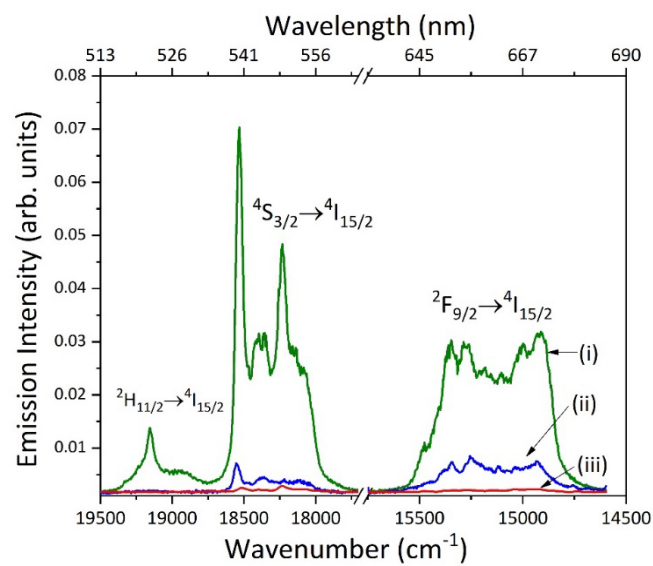

Fig. S5. Upconversion fluorescence spectra of $MF_2$:20 mol% $Yb^{3+}$/2 mol% $Er^{3+}$ nanoparticles where M=(i) $Sr^{2+}$, (ii) $Ca^{2+}$ and (iii) $Ba^{2+}$ under excitation at 980 nm using 188 mW laser power recorded at 10 K.